\documentstyle[12pt]{article}

\include{epsf}

\begin{document}

\thispagestyle{empty}

\begin{flushright}
UNIGRAZ-UTP-12-06-97 \\
hep-lat/9706010
\end{flushright}
\begin{center}
\vspace*{1cm}
{\Large Quantum Fluctuations versus Topology -\vskip3mm 
a Study in U(1)$_2$ Lattice Gauge Theory${}^*$  }
\vskip10mm
\centerline{ {\bf
C.R. Gattringer}}
\vskip 3mm
\centerline{Department of Physics and Astronomy,}
\centerline{University of British Columbia, Vancouver B.C., Canada}
\vskip 5mm
\centerline{ {\bf
I. Hip and C.B. Lang }}
\vskip3mm
\centerline{Institut f\"{u}r Theoretische Physik} 
\centerline{Universit\"at Graz, A-8010 Graz, Austria}
\end{center}
\vspace{10mm}
\begin{abstract}
Using the geometric definition of the topological charge we decompose
the path integral of 2-dimensional U(1) lattice gauge theory into
topological sectors.  In a Monte Carlo simulation we compute the
average value of the action as well as the distribution of its values
for each sector separately. These numbers are compared with analytic
lower bounds of the action which are relevant for classical
configurations carrying topological charge. We find that quantum
fluctuations entirely dominate the path integral. Our results for the
probability distribution of the Monte Carlo generated configurations
among the topological sectors can be understood by a
semi-phenomenological argument.
\end{abstract}
\bigskip \nopagebreak \begin{flushleft} \rule{2 in}{0.03cm}
\\ {\footnotesize \ 
${}^*$ Supported by Fonds zur F\"orderung der Wissenschaftlichen 
Forschung in \"Osterreich, Projects P11502-PHY and J01185-PHY.}
\end{flushleft}
\newpage
\setcounter{page}{1}
\noindent
The last few years have seen an ever increasing interest in studying
topological ideas in lattice gauge theories (see   
\cite{Ga96} for a recent review).
Many of these studies were motivated by computing statistical 
properties of instantons such as their average size and 
density distribution which are of interest for semiclassical models 
\cite{SchSh96} (see \cite{Sm92} for an analysis of these ideas for
the case of gauge group U(1) and 2 dimensions). Rather than using the 
lattice to extract information on some underlying classical 
configurations, in a series of articles \cite{GaHiLa97} 
we directly investigate possible implications of the 
existence of a topological structure for the lattice path integral. 

Here we study the characteristic values for the action 
of quantum fluctuations and compare them to analytic lower bounds 
of the action in each topological sector. Classical configurations
carrying topological charge typically have action in the vicinity 
of these bounds. It is interesting to analyze whether after adding 
quantum fluctuations the structure of the classical configurations is
still relevant for the physical picture.

The model under consideration is U(1) 
lattice gauge theory in 2 dimensions. Due to the simplicity of this model 
good statistics is easy to obtain. The model has the further advantage of 
possessing a simple form of the topological charge; the geometric
definition based on L\"uscher's work
\cite{Lu82} can be computed easily without making use of 
sophisticated techniques such as cooling or analyzing the spectrum
of the Dirac operator. These two facts allow to explore the physical 
mechanisms governing the role of topology on the lattice in a rather 
playful way.

Before we start developing our results we find it useful to comment on the
relevance of topological arguments in the continuum and on the lattice. In 
the continuum it is possible to classify classical, i.e.~differentiable
configurations with respect to their topological charge. It is 
crucial to note that these classical configurations are of 
measure zero in the continuum path integral \cite{CoLa73}. Thus 
in the continuum topological arguments cannot go beyond a 
semiclassical analysis. 
On the lattice the situation is different. In particular for the 
case of U(1)$_2$ lattice gauge theory one can assign a topological charge 
to all configurations (except for so-called exceptional configurations which 
are of measure zero). For SU(N)$_4$ it is believed, that the path integral 
in the continuum limit is dominated by smooth configurations which also can 
be assigned a topological charge. Thus in a certain sense, the lattice 
formulation is more powerful for studying topological ideas and the
lattice path integral can indeed be decomposed into topological sectors.
In particular the outlined questions can be formulated in a meaningful
way and a consistent physical picture can be obtained.
\\

We work on a 2-dimensional square lattice $\Lambda$ with length $L$. 
Lattice sites are denoted
as $x = (x_1,x_2)$ with $x_i = 1,2,...,L$. The lattice spacing is set
equal to 1. The gauge fields
are group elements $U_\mu(x) \in \mbox{U(1)}$ assigned to the 
links between nearest neighbours $x,x+\hat{\mu}$ and their action is given by
\begin{equation}
S \; = \; \beta \sum_{x \in \Lambda} \;
\Big[ \; 1  \; - \; \mbox{Re} \; U_P(x) \; \Big] \; .
\label{latact}
\end{equation}
The plaquette element is defined as 
$U_P(x) = U_1(x) U_2(x+\hat{1}) \overline{ U_1(x+\hat{2})} \; \overline{U_2(x)}$.
The gauge fields obey periodic boundary conditions $U_\mu(L+1,x_2) = 
U_\mu(1,x_2)$, $U_\mu(x_1,L+1) = U_\mu (x_1,1)$.

The model was simulated with the hybrid Monte Carlo 
algorithm\cite{DuKePe87} (this calculation is part of a another
project including also fermions \cite{GaHiLa97}) 
with 10-step trajectories adjusted
such that the acceptance rate in the Monte Carlo step was 0.8
in the mean. For all lattices sizes (up to $16\times 16$) we averaged
$10^5$ measurements. The individual measurements have been separated by 
10 updates (for small values of $\beta$) up to
500 updates (for $\beta=6$) in order to reduce correlations.

To check our algorithm we compare our results for simple observables
such as $\langle S/L^2 \rangle$ with the outcome of the analytic solution
for the model with open boundary conditions (see e.g. \cite{Zi96}). 
We find excellent agreement (within the small statistical errors), 
and only for the $L = 4$ lattice we 
observed small deviations due to the different boundary conditions. 

For the computation of the topological charge we use 
the geometric definition which is based on L\"uscher's 
idea of associating a principal
bundle to each lattice configuration and defining its topological charge
through the topological charge of the bundle 
\cite{Lu82}. The case of QED$_2$ was worked out
in \cite{lattop2}. One obtains
\begin{equation}
\nu[U] \; = \; \frac{1}{2\pi} \sum_{x \in \Lambda} \theta_P(x) \; ,
\label{toplat}
\end{equation}
where the plaquette angle $\theta_P(x)$ is given by 
$\theta_P(x) = -i \ln U_P(x)$
and restricted to the principal branch $\theta_P(x) \in (-\pi,\pi)$.
Note that configurations where $U_P(x) = -1$ for some $x$ are
so-called exceptional configurations and L\"uscher's definition 
does not assign a
value $\nu[U]$ to them. Those configurations are of measure 
zero in the path integral. 
\\

\begin{figure}[htbp]
\epsfysize=2.3in
\epsfbox[0 164 500 385] {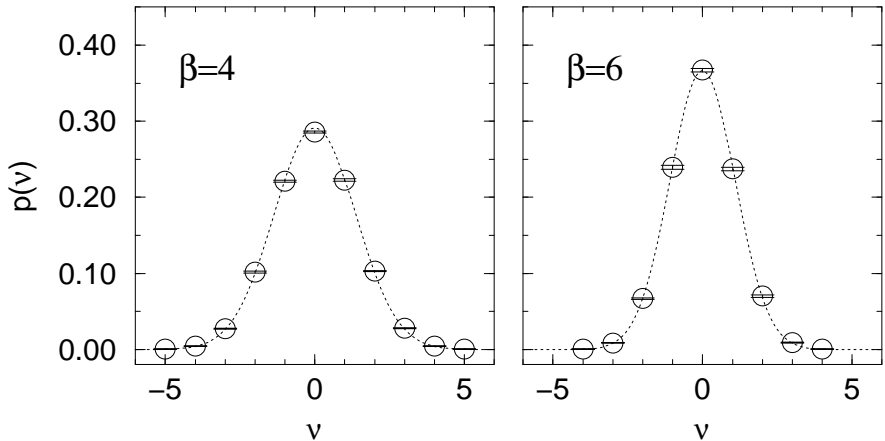}
\caption{
Probability distribution $p(\nu)$ of Monte Carlo generated 
configurations among
the topological sectors. The symbols give our results from 
a $16 \times 16$ lattice for $\beta = 4$ and $6$ ($10^5$ configurations 
each), the curve is the 
semi-phenomenological distribution function (\protect{\ref{spheno}}).
\label{confdist} }
\end{figure}

An interesting observable \cite{probdist,BaDuEiTh97} is the probability
distribution\footnote{ A physically more relevant observable is the
topological susceptibility which is essentially the inverse of the width
of the distribution in Fig. 1.  For our purposes however, the whole
distribution is more convenient since it contains more information than
the susceptibility.} of the Monte Carlo generated configurations among
the topological sectors.  In Fig.~\ref{confdist} we show our results
for a $16 \times 16$ lattice and $\beta = 4$ and 6.  For both values of
$\beta$ we observe a symmetric, Gaussian-like distribution centered at
$\nu = 0$. It becomes more peaked with increasing $\beta$.

Naturally the following question arises:
{\sl What is the mechanism behind this distribution of Monte Carlo 
configurations among the topological sectors?}
A first guess is, that there is a lower bound for the gauge field
action in each topological sector. Such a bound would force the action 
to higher values when increasing $|\nu|$ and through the Boltzmann factor 
lead to a suppression of configurations in higher sectors.

Here we use two different lower bounds, a strict bound and
a bound which holds only for sufficiently large $\beta$ but is
more interesting from a physical point of view.
The first one resembles the famous result \cite{BePoSchTy75}
for classical Yang-Mills configurations in 4 dimensions. 
It also can be derived using the Schwartz inequality \cite{Ra82}.
Essentially the same arguments can be repeated on the lattice.
We assume that the lattice gauge configuration $U$ is non-exceptional 
such that a
topological charge (\ref{toplat}) can be assigned. We use the
Schwartz inequality to obtain
\[
\left| \sum_{x \in \Lambda} \theta_P(x) \right|^2 \! =  
\left| \sum_{x \in \Lambda} \frac{\theta_P(x)}{1\!-\!e^{i\theta_P(x)}}
[1\!-\! e^{i\theta_P(x)}] \right|^2 \!
\leq \sum_{x \in \Lambda} [1\!- \cos\theta_{\!P }(x) ] 
\times \! \sum_{y \in \Lambda} 
\frac{\theta_P(y)^2}{1\!-\! \cos\theta_{\! P}(y)} .
\]
Using (\ref{latact}) and  (\ref{toplat}) one ends up with
\begin{equation}
S[U] \; \; \; \geq \; \; \; \frac{2 \pi^2}{L^2} \beta 
\; \Big| \nu [A] \Big|^2 \; D[U] \; \; \; \geq \; \; \;
\frac{2 \pi^2}{L^2} \beta 
\; \Big| \nu [A] \Big|^2 \; \frac{4}{\pi^2} \; ,
\label{lobolat}
\end{equation}
where
\begin{equation}
\frac{4}{\pi^2} \; \; \leq \; \; D[U] \; = \; 
2 \left[ \frac{1}{L^2} \sum_{x \in  \Lambda} 
\frac{\theta_P(x)^2}{1-\cos\theta_P(x)} \right]^{-1} \; \; \leq \; \; 1 \; .
\end{equation}
$D[U]$ is a measure for how much the average plaquette angle differs
from $0$. This functional takes values between $4/\pi^2$ and 1. In the 
latter case which corresponds to $\beta \rightarrow \infty$,
(\ref{lobolat}) equals the bound valid for classical configurations on a continuous 2-dimensional compact manifold
(replace $L^2$ by the volume of the manifold).

Another formula also serves as a lower bound for
sufficiently smooth configurations and is more stringent.
Using the fact that the function $(1 - \cos\theta)$ (the 
contribution of one plaquette to the gauge field action) is convex
for $|\theta| \leq \pi/2$ one finds
\[
\frac{1}{L^2} \sum_{x \in \Lambda} \; [ \; 1 - \cos\theta_P(x) \; ]
\; \; \geq \; \; 1 - \cos \left( \frac{1}{L^2} \sum_{x \in \Lambda}
\theta_P(x) \right) \; ,
\]
which implies
\begin{equation}
S[U] \; \; \geq \; \; \beta L^2 
\Big[ 1 - \cos \Big( \frac{2\pi}{L^2} \nu[U] \Big) \Big] \; .
\label{lobolat2}
\end{equation}
It has to be stressed, that this bound holds only if for 
all $x \in \Lambda$ the condition $|\theta_P(x)| \leq \pi/2$ is
fulfilled. However for large enough $\beta$ this is the case and
(\ref{lobolat2}) holds. For small $|\nu|$ it is more stringent 
than the exact bound (\ref{lobolat}) (compare Table 1). 

It is interesting to note, that the right hand side of (\ref{lobolat2})
is the value of the lattice action (\ref{latact}) for configurations
which correspond to continuum fields with constant electric field 
$2 \pi \nu / V$, where $V$ is the volume of the base manifold. In 
\cite{Jo90} the path integral for the Schwinger model an a continuous 
torus was constructed essentially using those constant field configurations
and adding quantum fluctuations. The lower bound (\ref{lobolat2}) is thus
of particular interest for analyzing the emergence of the physical picture
in the continuum \cite{Sm92,Jo90} from a lattice formulation.

\begin{table}
\begin{center}
\begin{tabular}{|l|c|c|c|c|c|}
\hline
parameters  & $\nu$ & $\overline{S}(\nu)/L^2$ & 
$\overline{S}_{ph}(\nu)/L^2$ & bound (\ref{lobolat}) & 
bound (\ref{lobolat2}) \\
\hline 
$ L = 8$ & 0 & 0.536(5) & 0.536 & 0 & 0 \\
$\beta = 4$ & 1 & 0.556(8) & 0.553 & 0.0078 & 0.0193 \\
 & 2 & 0.62(4) & 0.603 & 0.0312 & 0.0768 \\
\hline
$ L = 16$ & 0 & 0.543(3) & 0.543 & 0 & 0 \\
$\beta = 4$ & 1 & 0.545(3) & 0.544 & 0.0004 & 0.0012 \\
 & 2 & 0.549(4) & 0.547 & 0.0019 & 0.0048 \\
 & 3 & 0.555(9) & 0.552 & 0.0044 & 0.0108 \\
\hline
$ L = 16$ & 0 & 0.523(3) & 0.523 & 0 & 0 \\
$\beta = 6$ & 1 & 0.525(3) & 0.525 & 0.0007 & 0.0018 \\
 & 2 & 0.530(6) & 0.530 & 0.0029 & 0.0072 \\
 & 3 & 0.540(16) & 0.538 & 0.0066 & 0.0163 \\ 
\hline
\end{tabular}
\end{center}
\caption{Data for the action in various topological sectors. For several
values of $L, \beta$ and $\nu$ we give the average value of the action per 
plaquette $\overline{S}(\nu)/L^2$ the phenomenological value 
$\overline{S}_{ph}(\nu)/L^2$ from 
(\protect{\ref{gauss}}) and the values for the lower bounds
(\protect{\ref{lobolat}})
and (\protect{\ref{lobolat2}}) (also normalized with $L^{-2}$).}
\end{table}

\begin{figure}[htbp]
\begin{minipage}[t]{2.5in}
\epsfysize=1.96in
\epsfbox[29 73 461 390] {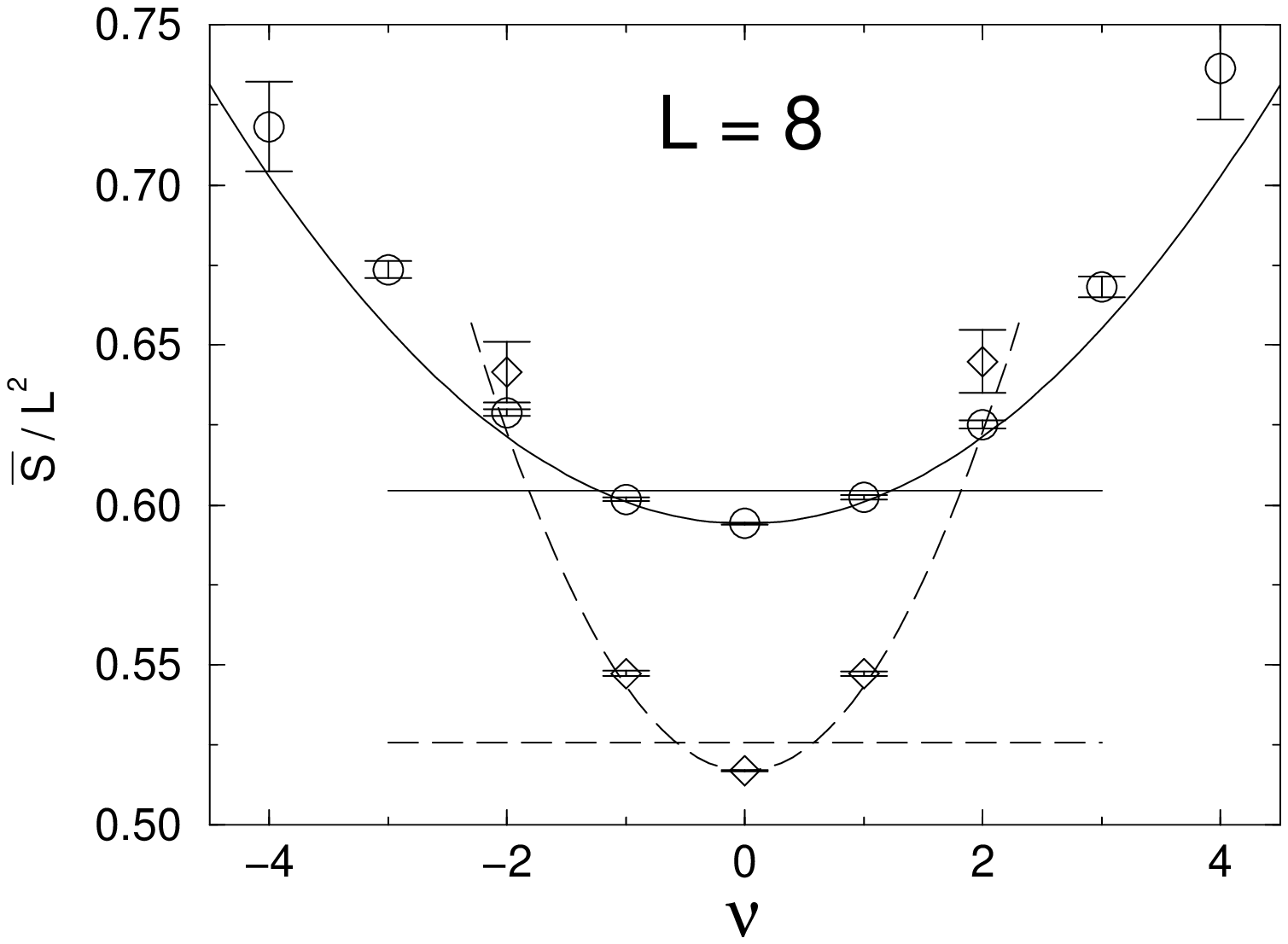}
\end{minipage}
\hfill
\begin{minipage}[t]{2.5in}
\epsfysize=1.96in
\epsfbox[29 73 461 390] {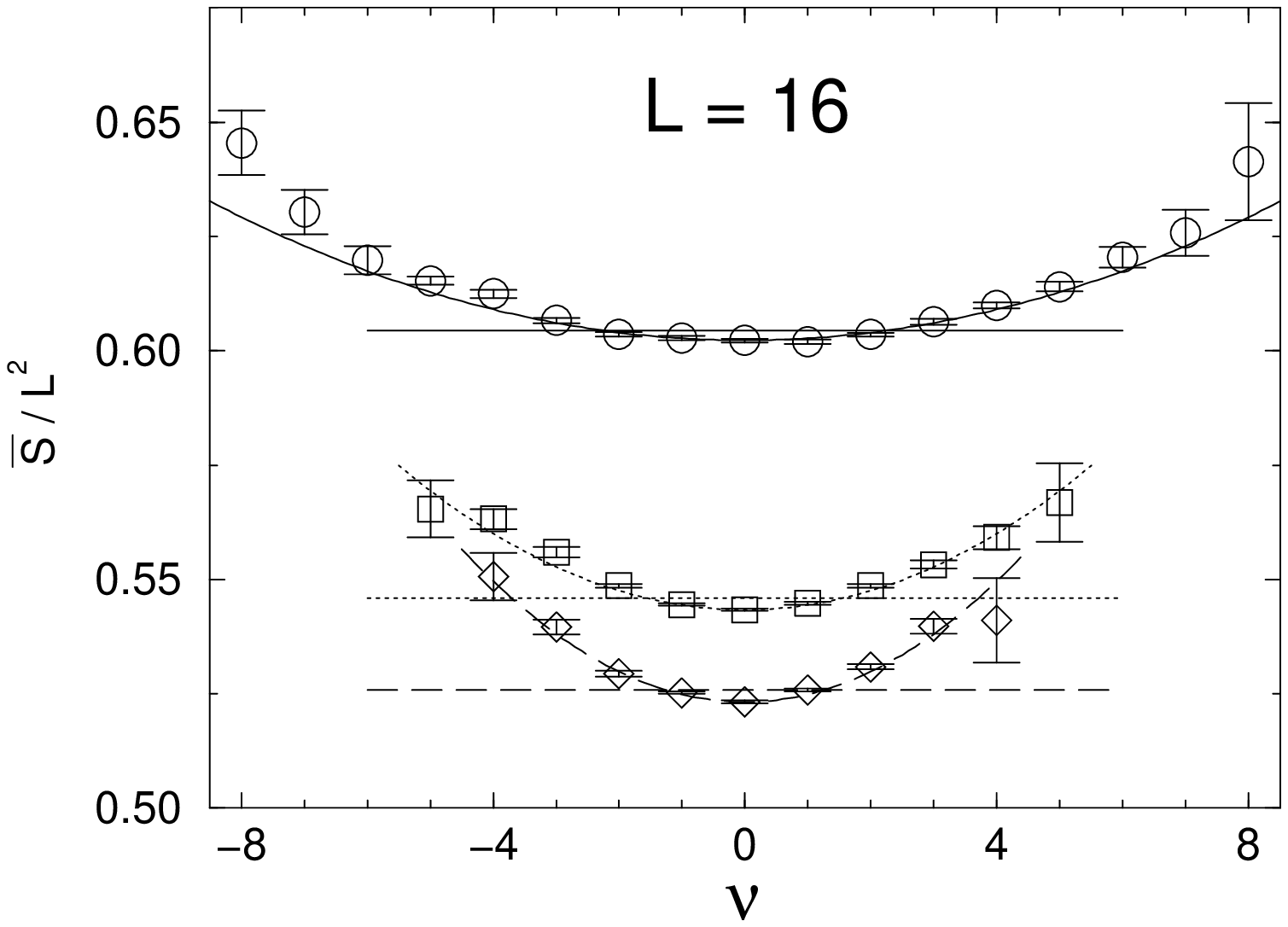}
\end{minipage}
\caption{
Average value $\overline{S}(\nu)/L^2$ of the action per plaquette 
in various topological sectors.
The symbols are the results of our Monte-Carlo simulation, the horizontal
lines are the values from the analytic solution for open boundary
conditions $[7]$. The curves connecting
the symbols come from the semi-phenomenological formula 
(\protect{\ref{gauss}}) to be discussed below. We show
our results for $L=8, 16$ and $\beta = 2$ (circles),
$\beta = 4$ (squares) and $\beta = 6$ (diamonds). \label{saver}}
\end{figure}

In order to study the outlined idea that lower bounds of the action in each 
sector are responsible for the distribution in Fig.~\ref{confdist},
we compute the average value $\overline{S}(\nu)/L^2$ of the action 
for each topological sector 
separately and compare this average with the values of the lower 
bounds (\ref{lobolat}) and (\ref{lobolat2}). 
Fig.~\ref{saver} and Table 1 give our results. 

In the figure
the symbols show the average value of the action per plaquette and 
the horizontal lines give the value of the analytic result for open 
boundary conditions \cite{Zi96} which is a
sum over all sectors. The curve connecting the 
symbols comes from the phenomenological formula (\ref{spheno}) below.
The average value of the action increases with increasing values of $|\nu|$
and thus the distribution of Fig.~\ref{confdist} can be understood
through the suppression by the Boltzmann factor.

What is indeed surprising is the fact, that the values of the
analytic lower bounds (\ref{lobolat}) and (\ref{lobolat2}) are far 
below the actual average values of the action in the path integral
(see Table 1). For all sectors the action stays close to the analytic 
value \cite{Zi96} which is  a sum
over all sectors (horizontal lines). Only the center of the distribution 
slightly dips below this value. The range of values for the analytic
lower bounds (and thus the action for the constant field configurations
\cite{Jo90}) is typically one order of magnitude smaller (see Table 1).
\\

\begin{figure}[htbp]
\epsfysize=2in
\epsfbox[4 471 484 652] {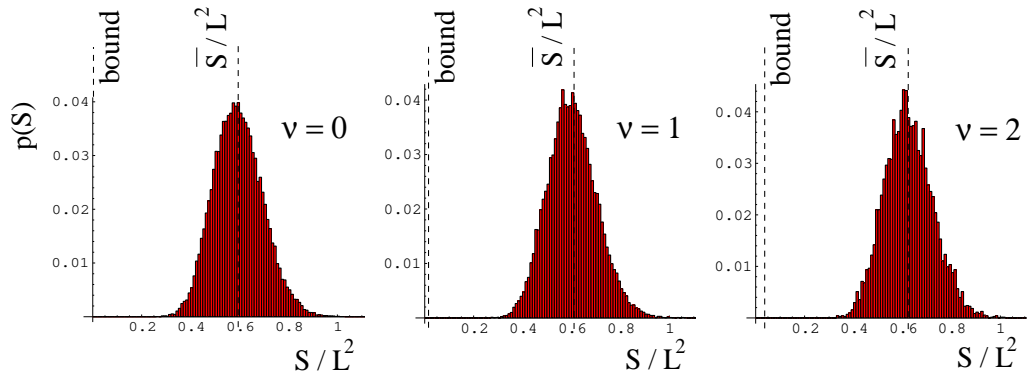}
\caption{
Probability distribution $p(S)$ 
of the action per plaquette in the sectors $\nu = 0, 1, 2$.
The data was taken on a $8\times 8$ lattice at $\beta = 2$.
The dotted vertical lines indicate the value of the lower bound
$(5)$ and the average $\overline{S}(\nu)/L^2$ for each sector. 
\label{histplot}}
\end{figure}

To study the quantum fluctuations further we analyze the distribution of the 
values of the action in a fixed sector. It can be computed by binning
the range of values of the action and counting the number of
configurations in each bin. In Fig.~\ref{histplot} we show histograms for
the probability distribution of the action per plaquette in the sectors with 
$\nu = 0,1$ and 2. The vertical lines show the average value 
$\overline{S}(\nu)$ of the action in each sector as well as the lower 
bound (\ref{lobolat2}).

It is obvious, that quantum fluctuations keep a large portion of the
configurations high above the lower bounds (\ref{lobolat}) and
(\ref{lobolat2}). For all sectors the configurations have action close
to the value given by the analytic solution \cite{Zi96} for open
boundary conditions. Comparison with the lower bound (\ref{lobolat2})
which gives the characteristic action for classical, topologically
non-trivial configurations shows the importance of quantum
fluctuations.  This dominance of the quantum fluctuations was already
conjectured in a study of the model on the continuous torus
\cite{Sm92}. We remark that we obtain similar distributions for other
values of $L$ and $\beta$. For increasing $L$ the total action of
the configurations tends to be more concentrated around the peak, and
the whole distribution is narrower, as expected. The
lower bounds become even less important. Increasing $\beta$ makes the
distribution less symmetric by increasing the weight of smaller values
of the action, essentially without altering the support of the
distribution.
\\

Although the analytic lower bounds do not seem to govern
the distribution of the values of the action in the 
fully quantized theory, it is still possible to use topological 
arguments in a phenomenological way to understand the distributions 
in Fig.~\ref{saver}. From the definition of the topological charge 
(\ref{toplat}) it follows that the average increase 
of the plaquette angle when increasing $|\nu|$ by 1
is given by $2 \pi/L^2$. 
Assuming that the effect of going from the trivial sector to a 
nontrivial sector can be taken into account by adding this average 
value to all plaquette angles one finds ($\theta^0_P(x)$ are the plaquette 
angles of a configuration in the trivial sector) 
$$
S_{ph}(\nu) = \beta \sum_{x \in \Lambda} \Big[ 1 -
\cos\Big(\theta^0_P(x) + \nu \frac{2\pi}{L^2} \Big) \Big] \; =
$$
$$
S(0)
+ \frac{2\,\pi\,\beta\,\nu}{L^2}\sum_{x \in \Lambda}\sin\theta^0_P(x)
+ \frac{2\,\pi^2\,\nu^2 }{L^4}\left[ - S(0) + \beta L^2 \right]
+ O(\nu^4/L^6) \;.
$$
In the second step we assumed that $|\nu|$ is small compared to $L^2$,
and expanded the cosine in $\nu / L^2$. The second order term has been
re-expressed in terms of the total action. 

The sum over  
$\sin\theta^0_P(x)$ may be neglected for two reasons.
On one hand in the sum over all configurations it vanishes
due to the symmetry of the Wilson action in the plaquette angle. Furthermore
(at least in order $\theta^0_P(x)$) it is proportional to the 
topological charge which is zero per ansatz. 
We end up with the following formula for the average value 
of the action
\begin{equation}
\overline{S}_{ph}(\nu) \; = \; 
\overline{S}(0) + \nu^2  \frac{2 \pi^2}{L^4}
\Big[ - \overline{S}(0) + \beta L^2 \Big] + O(\nu^4/L^6) \; .
\label{spheno}
\end{equation}
In Table 1 we give the values of (\ref{spheno}) for the corresponding
$\beta, L$ and $\nu$. The curves
in Fig.~\ref{saver} were drawn using (\ref{spheno}).
The formula gives reasonable values for the average 
action over a wide range of $\beta, L$ and $\nu$ with a slight tendency
to underestimate the Monte Carlo results for smaller $\beta$. 
This tendency can be understood
by the fact that $2\pi/ L^2$ only gives the minimal increase of the
average plaquette angle compatible with an increase of $|\nu|$ by 1.
Quantum fluctuations seem to lead to slightly 
higher values of the average plaquette angle and thus of 
the action. When increasing $\beta$ the fluctuations are more 
suppressed and the quality of (\ref{spheno}) increases.

We finally remark, that (\ref{spheno}) can be used to understand the
shape of the distribution in Fig. \ref{confdist} and to test the 
ergodicity of the updating algorithm. Raising (\ref{spheno}) to the
exponent and normalizing the result 
gives the Gaussian distribution
\begin{equation}
p(\nu) = N \exp \left( - \frac{1}{2}\, C \,\nu^2 \right) 
\; , \quad
C = \frac{4 \,\pi^2}{L^4}
\left[ - \overline{S}(0) + \beta \,L^2 \right] \; , 
\label{gauss}
\end{equation}
with normalization (note that $\nu$ is an integer)
$N^{-1} = \sum_{\nu} \exp( -\nu^2\, C/ 2 )$.
It should 
reproduce the probability distribution of the configurations among
the topological sectors. The dotted lines in Fig.~\ref{confdist}
show the Gaussian (8), and we find good agreement with the Monte Carlo 
data. The quality increases with $\beta$ and $L$ where formulas (\ref{spheno})
and (\ref{gauss}) are more reliable. Thus the probability distribution of Fig.~\ref{confdist} can indeed be understood using topological arguments.
We conclude that our updating algorithm is ergodic.

Using a different definition of the topological charge, Bardeen et al.
\cite{BaDuEiTh97} derive a similar formula for $p(\nu)$ from the
analytic solution \cite{Zi96}. In the limit $\beta \rightarrow \infty$,
where the two definitions of the topological charge are expected to
coincide, their result approaches our formula (\ref{gauss}). In fact,
replacing $\overline{S}(0)$ in (\ref{spheno}) by the analytic result
for infinite lattices, proportional to $[1-I_1(\beta)/I_0(\beta)]$
(which, however, sums all topological sectors) one
recovers the distribution given in \cite{BaDuEiTh97}.
\\

The most remarkable outcome of this study is the fact, that quantum 
fluctuations entirely dominate the path integral. At the values of 
$L$ and $\beta$ where our simulations were performed, the configurations
which essentially contribute to the path integral have values
of the action which are typically one order of magnitude larger than the
analytic lower bounds. This dominance of quantum fluctuations was already
conjectured for the model on the continuous torus in \cite{Sm92}.  
In the underlying construction \cite{Jo90}
the path integral is essentially decomposed into constant field configurations
that carry topological charge and saturate the lower bound of the action,
plus a field describing the quantum fluctuations.  
The observed importance of the quantum fluctuations shows that trying
to establish a physical picture where dominant classical configurations 
are used to understand the interplay of topological
charge and vacuum expectation values is questionable. 
 
Using a semi-phenomenological ansatz we were able to obtain a qualitative,
and in the $\beta \rightarrow \infty$ limit even quantitative 
understanding of the average behaviour of the action. This formula 
explains the probability distribution of Fig. \ref{confdist} in terms
of a topological argument and can  
be used to check the ergodicity of the updating algorithm.  

We believe, that although the technical problems are
much more severe, a similar study can -- in principle -- 
also be accomplished for lattice Yang Mills theory
in 4 dimensions. The possible outcome would be a better understanding 
of the relation between quantum fluctuations and classical configurations
carrying topological information. 
\\

We thank Philippe de Forcrand and Helmut Gausterer
for interesting discussions. C.R.~G. 
has also profited from remarks by Erhard Seiler and Gordon Semenoff.

\end{document}